# Limits on Anomalous $WW\gamma$ Couplings from $p\bar{p} \rightarrow W\gamma + X$ Events at $\sqrt{s} = 1.8$ TeV

S. Abachi,[14] B. Abbott,[28] M. Abolins,[25] B.S. Acharya,[43] I. Adam,[12] D.L. Adams,[37]
M. Adams,[17] S. Ahn,[14] H. Aihara,[22] G. Álvarez,[18] G.A. Alves,[10] E. Amidi,[29] N. Amos,[24]
E.W. Anderson,[19] S.H. Aronson,[4] R. Astur,[42] M.M. Baarmand,[42] A. Baden,[23]
V. Balamurali,[32] J. Balderston,[16] B. Baldin,[14] S. Banerjee,[43] J. Bantly,[5] J.F. Bartlett,[14]
K. Bazizi,[39] A. Belyaev,[26] J. Bendich,[22] S.B. Beri,[34] I. Bertram,[31] V.A. Bezzubov,[35]
P.C. Bhat,[14] V. Bhatnagar,[34] M. Bhattacharjee,[13] A. Bischoff,[9] N. Biswas,[32] G. Blazey,[30]
S. Blessing,[15] P. Bloom,[7] A. Boehnlein,[14] N.I. Bojko,[35] F. Borcherding,[14] J. Borders,[39]
C. Boswell,[9] A. Brandt,[14] R. Brock,[25] A. Bross,[14] D. Buchholz,[31] V.S. Burtovoi,[35]
J.M. Butler,[3] W. Carvalho,[10] D. Casey,[39] H. Castilla-Valdez,[11] D. Chakraborty,[42]
S.-M. Chang,[29] S.V. Chekulaev,[35] L.-P. Chen,[22] W. Chen,[42] S. Choi,[41] S. Chopra,[24]
B.C. Choudhary,[9] J.H. Christenson,[14] M. Chung,[17] D. Claes,[42] A.R. Clark,[22]
W.G. Cobau,[23] J. Cochran,[9] W.E. Cooper,[14] C. Cretsinger,[39] D. Cullen-Vidal,[5]
M.A.C. Cummings,[16] D. Cutts,[5] O.I. Dahl,[22] K. De,[44] K. Del Signore,[24] M. Demarteau,[14]
D. Denisov,[14] S.P. Denisov,[35] H.T. Diehl,[14] M. Diesburg,[14] G. Di Loreto,[25] P. Draper,[44]
J. Drinkard,[8] Y. Ducros,[40] L.V. Dudko,[26] S.R. Dugad,[43] D. Edmunds,[25] J. Ellison,[9]
V.D. Elvira,[42] R. Engelmann,[42] S. Eno,[23] G. Eppley,[37] P. Ermolov,[26] O.V. Eroshin,[35]
V.N. Evdokimov,[35] S. Fahey,[25] T. Fahland,[5] M. Fatyga,[4] M.K. Fatyga,[39] J. Featherly,[4]
S. Feher,[14] D. Fein,[2] T. Ferbel,[39] G. Finocchiaro,[42] H.E. Fisk,[14] Y. Fisyak,[7] E. Flattum,[25]
G.E. Forden,[2] M. Fortner,[30] K.C. Frame,[25] P. Franzini,[12] S. Fuess,[14] E. Gallas,[44]
A.N. Galyaev,[35] P. Gartung,[9] T.L. Geld,[25] R.J. Genik II,[25] K. Genser,[14] C.E. Gerber,[14]
B. Gibbard,[4] V. Glebov,[39] S. Glenn,[7] B. Gobbi,[31] M. Goforth,[15] A. Goldschmidt,[22]
B. Gómez,[1] G. Gomez,[23] P.I. Goncharov,[35] J.L. González Solís,[11] H. Gordon,[4] L.T. Goss,[45]
A. Goussiou,[42] N. Graf,[4] P.D. Grannis,[42] D.R. Green,[14] J. Green,[30] H. Greenlee,[14]
G. Griffin,[8] G. Grim,[7] N. Grossman,[14] P. Grudberg,[22] S. Grünendahl,[39] G. Guglielmo,[33]
J.A. Guida,[2] J.M. Guida,[5] W. Guryn,[4] S.N. Gurzhiev,[35] P. Gutierrez,[33] Y.E. Gutnikov,[35]
N.J. Hadley,[23] H. Haggerty,[14] S. Hagopian,[15] V. Hagopian,[15] K.S. Hahn,[39] R.E. Hall,[8]
S. Hansen,[14] J.M. Hauptman,[19] D. Hedin,[30] A.P. Heinson,[9] U. Heintz,[14]
R. Hernández-Montoya,[11] T. Heuring,[15] R. Hirosky,[15] J.D. Hobbs,[14] B. Hoeneisen,[1,†]
J.S. Hoftun,[5] F. Hsieh,[24] Ting Hu,[42] Tong Hu,[18] T. Huehn,[9] A.S. Ito,[14] E. James,[2]
J. Jaques,[32] S.A. Jerger,[25] J.Z.-Y. Jiang,[42] T. Joffe-Minor,[31] K. Johns,[2] M. Johnson,[14]
A. Jonckheere,[14] M. Jones,[16] H. Jöstlein,[14] S.Y. Jun,[31] C.K. Jung,[42] S. Kahn,[4]
G. Kalbfleisch,[33] J.S. Kang,[20] R. Kehoe,[32] M.L. Kelly,[32] L. Kerth,[22] C.L. Kim,[20]
S.K. Kim,[41] A. Klatchko,[15] B. Klima,[14] B.I. Klochkov,[35] C. Klopfenstein,[7] V.I. Klyukhin,[35]
V.I. Kochetkov,[35] J.M. Kohli,[34] D. Koltick,[36] A.V. Kostritskiy,[35] J. Kotcher,[4]
A.V. Kotwal,[12] J. Kourlas,[28] A.V. Kozelov,[35] E.A. Kozlovski,[35] J. Krane,[27]
M.R. Krishnaswamy,[43] S. Krzywdzinski,[14] S. Kunori,[23] S. Lami,[42] H. Lan,[14,*]
G. Landsberg,[14] B. Lauer,[19] J-F. Lebrat,[40] A. Leflat,[26] H. Li,[42] J. Li,[44] Y.K. Li,[31]
Q.Z. Li-Demarteau,[14] J.G.R. Lima,[38] D. Lincoln,[24] S.L. Linn,[15] J. Linnemann,[25]

R. Lipton,[14] Q. Liu,[14,*] Y.C. Liu,[31] F. Lobkowicz,[39] S.C. Loken,[22] S. Lökös,[42] L. Lueking,[14] A.L. Lyon,[23] A.K.A. Maciel,[10] R.J. Madaras,[22] R. Madden,[15] L. Magaña-Mendoza,[11] S. Mani,[7] H.S. Mao,[14,*] R. Markeloff,[30] L. Markosky,[2] T. Marshall,[18] M.I. Martin,[14] B. May,[31] A.A. Mayorov,[35] R. McCarthy,[42] J. McDonald,[15] T. McKibben,[17] J. McKinley,[25] T. McMahon,[33] H.L. Melanson,[14] K.W. Merritt,[14] H. Miettinen,[37] A. Mincer,[28] J.M. de Miranda,[10] C.S. Mishra,[14] N. Mokhov,[14] N.K. Mondal,[43] H.E. Montgomery,[14] P. Mooney,[1] H. da Motta,[10] M. Mudan,[28] C. Murphy,[17] F. Nang,[5] M. Narain,[14] V.S. Narasimham,[43] A. Narayanan,[2] H.A. Neal,[24] J.P. Negret,[1] P. Nemethy,[28] D. Nešić,[5] M. Nicola,[10] D. Norman,[45] L. Oesch,[24] V. Oguri,[38] E. Oltman,[22] N. Oshima,[14] D. Owen,[25] P. Padley,[37] M. Pang,[19] A. Para,[14] Y.M. Park,[21] R. Partridge,[5] N. Parua,[43] M. Paterno,[39] J. Perkins,[44] M. Peters,[16] H. Piekarz,[15] Y. Pischalnikov,[36] V.M. Podstavkov,[35] B.G. Pope,[25] H.B. Prosper,[15] S. Protopopescu,[4] D. Pušeljić,[22] J. Qian,[24] P.Z. Quintas,[14] R. Raja,[14] S. Rajagopalan,[42] O. Ramirez,[17] P.A. Rapidis,[14] L. Rasmussen,[42] S. Reucroft,[29] M. Rijssenbeek,[42] T. Rockwell,[25] N.A. Roe,[22] P. Rubinov,[31] R. Ruchti,[32] J. Rutherfoord,[2] A. Sánchez-Hernández,[11] A. Santoro,[10] L. Sawyer,[44] R.D. Schamberger,[42] H. Schellman,[31] J. Sculli,[28] E. Shabalina,[26] C. Shaffer,[15] H.C. Shankar,[43] R.K. Shivpuri,[13] M. Shupe,[2] H. Singh,[34] J.B. Singh,[34] V. Sirotenko,[30] W. Smart,[14] A. Smith,[2] R.P. Smith,[14] R. Snihur,[31] G.R. Snow,[27] J. Snow,[33] S. Snyder,[4] J. Solomon,[17] P.M. Sood,[34] M. Sosebee,[44] N. Sotnikova,[26] M. Souza,[10] A.L. Spadafora,[22] R.W. Stephens,[44] M.L. Stevenson,[22] D. Stewart,[24] D.A. Stoianova,[35] D. Stoker,[8] K. Streets,[28] M. Strovink,[22] A. Sznajder,[10] P. Tamburello,[23] J. Tarazi,[8] M. Tartaglia,[14] T.L.T. Thomas,[31] J. Thompson,[23] T.G. Trippe,[22] P.M. Tuts,[12] N. Varelas,[25] E.W. Varnes,[22] D. Vititoe,[2] A.A. Volkov,[35] A.P. Vorobiev,[35] H.D. Wahl,[15] G. Wang,[15] J. Warchol,[32] G. Watts,[5] M. Wayne,[32] H. Weerts,[25] A. White,[44] J.T. White,[45] J.A. Wightman,[19] S. Willis,[30] S.J. Wimpenny,[9] J.V.D. Wirjawan,[45] J. Womersley,[14] E. Won,[39] D.R. Wood,[29] H. Xu,[5] R. Yamada,[14] P. Yamin,[4] C. Yanagisawa,[42] J. Yang,[28] T. Yasuda,[29] P. Yepes,[37] C. Yoshikawa,[16] S. Youssef,[15] J. Yu,[14] Y. Yu,[41] Q. Zhu,[28] Z.H. Zhu,[39] D. Zieminska,[18] A. Zieminski,[18] E.G. Zverev,[26] and A. Zylberstejn[40]

(DØ Collaboration)

[1]*Universidad de los Andes, Bogotá, Colombia*
[2]*University of Arizona, Tucson, Arizona 85721*
[3]*Boston University, Boston, Massachusetts 02215*
[4]*Brookhaven National Laboratory, Upton, New York 11973*
[5]*Brown University, Providence, Rhode Island 02912*
[6]*Universidad de Buenos Aires, Buenos Aires, Argentina*
[7]*University of California, Davis, California 95616*
[8]*University of California, Irvine, California 92717*
[9]*University of California, Riverside, California 92521*
[10]*LAFEX, Centro Brasileiro de Pesquisas Físicas, Rio de Janeiro, Brazil*
[11]*CINVESTAV, Mexico City, Mexico*
[12]*Columbia University, New York, New York 10027*
[13]*Delhi University, Delhi, India 110007*
[14]*Fermi National Accelerator Laboratory, Batavia, Illinois 60510*

[15]*Florida State University, Tallahassee, Florida 32306*
[16]*University of Hawaii, Honolulu, Hawaii 96822*
[17]*University of Illinois at Chicago, Chicago, Illinois 60607*
[18]*Indiana University, Bloomington, Indiana 47405*
[19]*Iowa State University, Ames, Iowa 50011*
[20]*Korea University, Seoul, Korea*
[21]*Kyungsung University, Pusan, Korea*
[22]*Lawrence Berkeley National Laboratory and University of California, Berkeley, California 94720*
[23]*University of Maryland, College Park, Maryland 20742*
[24]*University of Michigan, Ann Arbor, Michigan 48109*
[25]*Michigan State University, East Lansing, Michigan 48824*
[26]*Moscow State University, Moscow, Russia*
[27]*University of Nebraska, Lincoln, Nebraska 68588*
[28]*New York University, New York, New York 10003*
[29]*Northeastern University, Boston, Massachusetts 02115*
[30]*Northern Illinois University, DeKalb, Illinois 60115*
[31]*Northwestern University, Evanston, Illinois 60208*
[32]*University of Notre Dame, Notre Dame, Indiana 46556*
[33]*University of Oklahoma, Norman, Oklahoma 73019*
[34]*University of Panjab, Chandigarh 16-00-14, India*
[35]*Institute for High Energy Physics, 142-284 Protvino, Russia*
[36]*Purdue University, West Lafayette, Indiana 47907*
[37]*Rice University, Houston, Texas 77005*
[38]*Universidade Estadual do Rio de Janeiro, Brazil*
[39]*University of Rochester, Rochester, New York 14627*
[40]*CEA, DAPNIA/Service de Physique des Particules, CE-SACLAY, France*
[41]*Seoul National University, Seoul, Korea*
[42]*State University of New York, Stony Brook, New York 11794*
[43]*Tata Institute of Fundamental Research, Colaba, Bombay 400005, India*
[44]*University of Texas, Arlington, Texas 76019*
[45]*Texas A&M University, College Station, Texas 77843*

## Abstract

We have measured the $WW\gamma$ gauge boson coupling parameters using $p\bar{p} \to \ell\nu\gamma + X$ ($\ell = e, \mu$) events at $\sqrt{s} = 1.8$ TeV. The data, corresponding to an integrated luminosity of 89.1 pb$^{-1}$, were collected using the DØ detector at the Fermilab Tevatron Collider. The measured cross section times branching ratio for $p\bar{p} \to W\gamma + X$ with $p_T^\gamma > 10$ GeV/c and $\mathcal{R}_{\ell\gamma} > 0.7$ is $11.8^{+1.7}_{-1.6} \pm 2.0$ pb, in agreement with the Standard Model prediction. The one degree of freedom 95% confidence level limits on individual $CP$-conserving parameters are $-0.98 < \Delta\kappa < 1.01$ and $-0.33 < \lambda < 0.31$. Similar limits are set on the $CP$-violating coupling parameters.

Measurement of the self-couplings of the gauge bosons provides important tests of the Standard Model (SM) of electroweak interactions. Recent limits on the $WW\gamma$ coupling parameters have been obtained by UA2 [1], CDF [2], DØ [3], and CLEO [4]. The hadron collider measurements relied on direct observation of $W\gamma$ final states, while the CLEO result used the observation of $b \rightarrow s\gamma$ decays.

The $WW\gamma$ coupling is fixed by the $SU(2)_L \otimes U(1)_Y$ symmetry of the SM. An effective Lagrangian [5] with four coupling parameters ($\kappa$, $\lambda$, $\tilde{\kappa}$ and $\tilde{\lambda}$) is introduced to allow for anomalies in the $WW\gamma$ interaction vertex. In the SM, the coupling parameters have the values $\Delta\kappa \equiv \kappa - 1 = 0$, $\lambda = \tilde{\lambda} = \tilde{\kappa} = 0$. $\Delta\kappa$ and $\lambda$ are related to the magnetic dipole and electric quadrupole moments of the $W$ boson. The non-Abelian nature of the SM manifests itself explicitly in the values of the coupling parameters; the minimal $U(1)_{EM}$ coupling of the photon to the electric charge of the $W$ boson would have the non-SM values of $\Delta\kappa = -1$ and $\lambda = 0$. The $\kappa$ and $\lambda$ terms are *CP*-invariant while the $\tilde{\kappa}$ and $\tilde{\lambda}$ terms violate *CP*. The pairs of *CP*-conserving and *CP*-violating couplings are considered independently, because they do not interfere with each other.

For non-SM couplings, the effective Lagrangian violates partial wave unitarity at high energies [5,6], so it is necessary to introduce form factors for each of the coupling parameters with a cut-off scale $\Lambda$. In this analysis, we assume dipole form factors of the type $\Delta\kappa(\hat{s}) = \Delta\kappa/(1+\hat{s}/\Lambda^2)^2$ where $\sqrt{\hat{s}}$ is the $W\gamma$ invariant mass and $\Lambda$ is the scale. We used $\Lambda = 1.5$ TeV in this analysis. Anomalies in the $WW\gamma$ interaction cause an increase in the total cross section for $p\bar{p} \rightarrow W\gamma + X$ and result in photons with higher transverse momentum ($p_T$) than those for the SM $WW\gamma$ interaction.

The analyses described here use $p\bar{p} \rightarrow \ell\nu\gamma + X$ ($\ell = e, \mu$) events observed with the DØ detector during the 1992–1993 and 1993–1995 runs of the Fermilab Tevatron Collider, corresponding to a total integrated luminosity of $89.1 \pm 9.4$ pb$^{-1}$. The DØ detector and data collection systems are described in Ref. [7]. Our analysis of events in 13.8 pb$^{-1}$ from the 1992–1993 Tevatron run has been described in an earlier paper [3]; this letter focuses on the details of the 1993–1995 analysis of $75.3 \pm 9.0$ pb$^{-1}$ and gives combined results from both analyses.

Events from $W \rightarrow e\nu$ decays were collected with a trigger that required missing transverse energy $\not\!\!E_T > 15$ GeV and an isolated electromagnetic (EM) cluster with transverse energy $E_T > 20$ GeV. The offline kinematic requirements imposed on this sample were: $E_T^e > 25$ GeV, $\not\!\!E_T > 25$ GeV, and $M_T(e, \not\!\!E_T) > 40$ GeV/c$^2$, where $M_T$ is the transverse mass $[2E_T^e \not\!\!E_T(1 - \cos\phi^{e\nu})]^{1/2}$ of the electron and $\not\!\!E_T$ vector separated by $\phi^{e\nu}$ in azimuth. The electron clusters were required to pass identical selection criteria, based on their shower profile and tracking information, as in our earlier analysis [3]. The electrons were required to have $|\eta| < 1.1$ in the central calorimeter (CC) or $1.5 < |\eta| < 2.5$ in the end calorimeters (EC), where $\eta$ is the pseudorapidity.

Events from the $W \rightarrow \mu\nu$ decay were collected with a trigger that required a muon with transverse momentum $p_T^\mu > 8$ GeV/c and an EM cluster with $E_T > 7$ GeV. The offline requirements imposed on this sample were: $|\eta^\mu| < 1.0$, $p_T^\mu > 15$ GeV/c, $\not\!\!E_T > 15$ GeV, and $M_T(\mu, \not\!\!E_T) > 30$ GeV/c$^2$. The quality cuts imposed on muons were similar to those used in the earlier analysis. Muon candidates were identified by a track traversing the muon proportional drift chambers and iron toroid magnet. They were required to match a charged track in the central drift chambers and to be isolated from nearby jets by at least 0.5 units in $\eta$-$\phi$ space.

Events in which a second muon was found in the muon chambers were rejected, as this is the signature of a $Z(\mu\mu)\gamma$ event. We also rejected events which contained an additional muon identified by an energy deposition in the longitudinally segmented calorimeter, up to $|\eta| < 2.7$, forming a track consistent with a muon and pointing to the interaction vertex. Any $W(\mu\nu)\gamma$ candidate event with a muon identified with calorimeter energy within $\phi^{\mu\nu} < 0.3$ radians of the missing transverse energy was rejected. This cut was found to be $(93 \pm 2)\%$ efficient for the $W(\mu\nu)\gamma$ signal, while accepting only $(35 \pm 3)\%$ of the $Z(\mu\mu)\gamma$ background.

The photons for both analyses were found in the same fiducial volume as the electrons, but with a lower kinematic requirement: $p_T^\gamma > 10$ GeV/c. We required the leptons and photons to be separated from each other by $\mathcal{R}_{\ell\gamma} > 0.7$ units in $\eta$-$\phi$ space. The selection criteria for photons are identical to those used in the earlier analysis, with requirements made on the electromagnetic shower profiles of the photons and the absence of an associated track in the drift chambers. Approximately 30% of the photons from $W\gamma$ events are expected to be found in the forward electromagnetic calorimeters.

In the electron channel, we rejected photon candidates which had unreconstructed tracks lying between the EM cluster and the event vertex. This cut was applied to reject backgrounds from processes (labelled $\ell eX$) which produced missing transverse energy, a high-$p_T$ lepton, and an electron with an unreconstructed track. These backgrounds are from $t\bar{t}$ and $WW$ pair production with a subsequent $W \to e\nu$ decay, and in the electron channel, high-$p_T$ $Z \to ee$ and QCD multijet production. The number of hits in the tracking chambers was counted in a road defined between the EM cluster and the event vertex. Photons were rejected if the number of hits exceeded a threshold defined separately for each of the tracking chambers. The efficiency and rejection of this quality cut were found by using $Z \to ee$ data. For the efficiency calulation we used the emulated photon technique in which roads pointing to the electron clusters were rotated in $\phi$ by $\pi/2$. The hit-counting cut was $(83 \pm 1)\%$ efficient for CC photons and $(70 \pm 3)\%$ efficient for EC photons, and rejected $(89 \pm 2)\%$ of both CC and EC electrons. In the $W(\mu\nu)\gamma$ analysis, the $\ell eX$ background was relatively small, so the hit-counting cut was not applied. The background from $\ell eX$ events in the $W(e\nu)\gamma$ channel was estimated from the number of events in the 1993–1995 data set which passed the same selection criteria as for the $W\gamma$ candidate events, but with the photon identification changed to require a track pointing to the EM cluster. We then multiplied this number by the measured efficiencies of the central tracking chambers, $(83.1 \pm 0.5)\%$ for central EM clusters and $(85.6 \pm 0.8)\%$ for forward EM clusters. Monte Carlo samples were used to estimate the $\ell eX$ background in the $W(\mu\nu)\gamma$ channel.

The background estimates and the total number of observed events for the two decay modes of the $W$ boson are summarized in Table I. The dominant background to $W\gamma$ events is from $W$ + jets processes in which a jet fragments into a neutral meson such as a $\pi^0$, which then decays to photons. The probability $\mathcal{P}$ for this to occur was estimated from a large data sample of multijet events. We found $\mathcal{P} \sim 11(13) \times 10^{-4}$ for CC (EC) photons, before applying the hit-counting cut. This additional cut, used in the $W(e\nu)\gamma$ analysis, reduces $\mathcal{P}$ by the measured efficiency of that cut. As in our earlier analysis, the $E_T$-dependent fraction of true photons in the multijet sample was applied as a correction to the measured values of $\mathcal{P}$, and introduced an uncertainty of 25% to the $W$ + jets background estimate. The backgrounds due to $Z\gamma$ and $W(\tau\nu)\gamma$ were estimated using the $Z\gamma$ event generator of Baur and Berger [8] and the ISAJET program [9], respectively, followed by a full detector simulation using the GEANT program [10].

The trigger and offline lepton selection efficiencies were estimated using $Z \to \ell\bar{\ell}$ and $W \to \ell\nu$ events. The trigger efficiencies were $(98\pm2)\%$ for the electron channel and $(71\pm3)\%$ for the muon channel. The offline selection efficiencies for electrons were $(77\pm1)\%$ in the CC and $(76\pm1)\%$ in the EC, while the muon selection efficiency was $(57\pm2)\%$. The detection efficiency for photons with $p_T > 25$ GeV/c was determined using electrons from $Z$ decays. For photons with a lower $p_T^\gamma$ there was a decrease in detection efficiency due to the cluster shape requirements. This decrease was estimated using Monte Carlo photons overlaid with minimum bias events from data, weighted to reflect the instantaneous luminosity profile of the 1993–1995 data. In the CC (EC), averages of approximately 10% (20%) of photons were also lost due to $e^+e^-$ pair conversions. The probability for tracks from other particles in the event to overlap the photon clusters was measured to be $(13.9\pm0.5)\%$ for CC photons and $(16.1\pm0.8)\%$ for EC photons. Combining these inefficiencies with the $p_T$-dependent photon detection efficiency, we estimated that the overall photon selection efficiency, before the introduction of the hit-counting cut, was $(45\pm4)\%$ in the CC and $(49\pm4)\%$ in the EC at $p_T^\gamma = 10$ GeV/c, and that it increased to $(71\pm7)\%$ in the CC and $(57\pm5)\%$ in the EC for $p_T^\gamma > 25$ GeV/c.

The kinematic and geometric acceptances were calculated as a function of coupling parameter values using the Monte Carlo program of Baur and Zeppenfeld [6], in which $W\gamma$ production and radiative decay processes are generated to leading order; higher order QCD effects are approximated by a K-factor of 1.335. We used the MRSD$-'$ parton distribution functions [11] and simulated the $p_T$ distribution of the $W\gamma$ system by using the observed $W$ $p_T$ spectrum in the inclusive $W \to e\nu$ data. The kinematic and fiducial acceptance for SM $W\gamma$ events in the DØ detector was $(11\pm1)\%$ for $W(e\nu)\gamma$ and $(19\pm1)\%$ for $W(\mu\nu)\gamma$.

The cross section times branching ratio $\sigma(p\bar{p} \to W\gamma+X)\times \mathrm{B}(W\to\ell\nu)$, where $\ell = e$ or $\mu$, was calculated for $p_T^\gamma > 10$ GeV/c and $\mathcal{R}_{\ell\gamma} > 0.7$ and found to be $13.1^{+3.2}_{-2.8}$ (stat)$\pm2.1$ (syst) $\pm$ 1.6 (lum) pb, where the first error is the $1\sigma$ uncertainty from Poisson statistics and the second term is the systematic error including the uncertainty in the $e/\mu/\gamma$ efficiencies and the uncertainty in the background estimates. The third error is due to the uncertainty in the calculation of the integrated luminosity. The observed cross section agrees with the SM prediction [6], $\sigma(p\bar{p} \to W\gamma+X)\times \mathrm{B}(W\to\ell\nu) = 12.5\pm1.0$ pb, where the uncertainty is due to the choice of parton distribution functions, the $Q^2$ scale at which the parton distribution functions are evaluated, and the $p_T$ distribution of the $W\gamma$ system.

Study of the individual leptonic decay modes of the $W$ boson can be considered as independent analyses with a common subset of systematic uncertainties. The same holds true for the data collected during the earlier Tevatron run. Combining the analyses from both runs, we observe 127 candidate events, with $84.1^{+12.3}_{-11.3}\pm8.7$ ascribed to signal. The first error is the $1\sigma$ uncertainty due to Poisson statistics and the second is due to the uncertainties in the background estimates. The number of signal events from each experiment is shown in Table I. The measured cross section from the combined sample of 84.1 signal events is $\sigma(p\bar{p} \to W\gamma+X)\times \mathrm{B}(W\to\ell\nu) = 11.8^{+1.7}_{-1.6}$ (stat) $\pm$ 1.6 (syst) $\pm$ 1.2 (lum) pb.

Figure 1 shows kinematic distributions of the 127 $W\gamma$ candidates from the combined data sets, along with the SM expectations and the background estimates. The spectrum showing the three-body transverse mass, $M_T(\ell,\gamma;\not{E}_T) = M_T(W,\gamma)$, is of particular interest.

$W\gamma$ events with a large $M_T(W,\gamma)$ have greater sensitivity to anomalous couplings whereas events with a three-body transverse mass below the mass of the $W$ boson are dominated by photons from radiative $W$ decays.

Limits on the anomalous coupling parameters are set by performing a binned maximum likelihood fit to the $p_T^\gamma$ spectrum for the individual decay modes of the $W$ boson. For each $p_T^\gamma$ spectrum, we calculate the probability for the sum of the background estimate and the Monte Carlo prediction to fluctuate to the observed number of events. The final limits on the anomalous coupling parameters are from a combined likelihood of both decay channels from this analysis and from our 1992–1993 analysis. For the 1993–1995 electron channel data, we imposed the requirement $M_T(W,\gamma) > 90$ GeV/c$^2$, which was found to increase the sensitivity to anomalous couplings by 10% for this data set alone. We did not impose this cut in the muon channel due to the less precise measurement of the muon momentum. The uncertainties in background estimates, efficiencies, acceptances, and integrated luminosity are folded into the combined likelihood function with Gaussian distributions. Uncertainties common to more than one analysis, e.g. theoretical uncertainties, are folded into the likelihood function only once.

The one- and two-dimensional [12] 95% confidence level (CL) contours for the *CP*-conserving parameters are shown in Fig. 2. The 95% CL limits for the individual coupling parameters, when all other parameters are held to their SM values, are: $-0.98 < \Delta\kappa < 1.01$, $-0.33 < \lambda < 0.31$, $-0.99 < \tilde{\kappa} < 1.00$, and $-0.32 < \tilde{\lambda} < 0.32$. For the *CP*-conserving couplings, the limits can be read from the one-dimensional 95% CL contour of Fig. 2. For example, the $\Delta\kappa$ limits correspond to the points of intersection of the inner ellipse with the $\lambda = 0$ axis. These results are the most stringent limits on anomalous $WW\gamma$ coupling parameters set by direct observation of $W\gamma$ events.

Assuming that the *CP*-violating couplings are zero, the $U(1)_{EM}$-only coupling ($\kappa = 0$, $\lambda = 0$) is excluded at the 86% CL. Making the further assumption that $\lambda = 0$, this point is exluded at the 95% CL. Exclusion of this point is direct evidence that the photon couples to more than just the electric charge of the $W$ boson.

We thank U. Baur for providing us with the $W\gamma$ and $Z\gamma$ Monte Carlo programs and for many helpful discussions. We thank the staffs at Fermilab and the collaborating institutions for their contributions to the success of this work, and acknowledge support from the Department of Energy and National Science Foundation (U.S.A.), Commissariat à L'Energie Atomique (France), Ministries for Atomic Energy and Science and Technology Policy (Russia), CNPq (Brazil), Departments of Atomic Energy and Science and Education (India), Colciencias (Colombia), CONACyT (Mexico), Ministry of Education and KOSEF (Korea), CONICET and UBACyT (Argentina), and the A.P. Sloan Foundation.

# REFERENCES

* Visitor from IHEP, Beijing, China.
† Visitor from Univ. San Francisco de Quito, Ecuador.

[1] UA2 Collaboration, J. Alitti *et al.*, Phys. Lett. **B277**, 194 (1992).
[2] CDF Collaboration, F. Abe *et al.*, Phys. Rev. Lett. **74**, 1936 (1995); **75**, 1017 (1995)
[3] DØ Collaboration, S. Abachi *et al.*, Phys. Rev. Lett. **75**, 1034 (1995); **75**, 1023 (1995); **77**, 3303 (1996)
[4] S. Stone and S. Playfer, Int. J. Mod. Phys. **A10**, 4107 (1995).
[5] K. Hagiwara, *et al.*, Nucl. Phys. **B282**, 253 (1987).
[6] U. Baur and D. Zeppenfeld, Nucl. Phys. **B308**, 127 (1988); U. Baur and E.L. Berger, Phys. Rev. D **41**, 1476 (1990).
[7] DØ Collaboration, S. Abachi *et al.*, Nucl. Instrum. Methods A **338**, 185 (1994).
[8] U. Baur and E.L. Berger, Phys. Rev. D **47**, 4889 (1993).
[9] F. Paige and S. Protopopescu, BNL Report BNL38034, 1986 (unpublished), release V6.49.
[10] F. Carminati *et al.*, "GEANT Users Guide," CERN Program Library, December 1991 (unpublished).
[11] A.D. Martin, W.J. Stirling and R.G. Roberts, Phys. Rev. D **47**, 867 (1993).
[12] The one- and two-dimensional 95% CL contours correspond to a 1.92 and 3.00 decrease in the likelihood function respectively.

## TABLES

TABLE I. Numbers of signal and background events.

| | 1992–1993 | | 1993–1995 | |
|---|---|---|---|---|
| | $e\nu\gamma$ | $\mu\nu\gamma$ | $e\nu\gamma$ | $\mu\nu\gamma$ |
| Luminosity | 13.8 $\text{pb}^{-1}$ | | 75.3 $\text{pb}^{-1}$ | |
| Backgrounds: | | | | |
| $W$ + jets | $1.7 \pm 0.9$ | $1.3 \pm 0.7$ | $11.5 \pm 2.3$ | $15.5 \pm 4.5$ |
| $Z\gamma$ | $0.1 \pm 0.1$ | $2.7 \pm 0.8$ | $0.4 \pm 0.1$ | $5.2 \pm 0.4$ |
| $W(\tau\nu)\gamma$ | $0.2 \pm 0.1$ | $0.4 \pm 0.1$ | $0.6 \pm 0.1$ | $1.7 \pm 0.3$ |
| $\ell eX$ | - | - | $0.7 \pm 0.1$ | $0.9 \pm 0.3$ |
| Total Bkgd | $2.0 \pm 0.9$ | $4.4 \pm 1.1$ | $13.2 \pm 2.3$ | $23.3 \pm 4.6$ |
| # Observed | 11 | 12 | 46 | 58 |
| Total Signal | $9.0^{+4.2}_{-3.1}$ | $7.6^{+4.4}_{-3.2}$ | $32.8^{+7.8}_{-6.8}$ | $34.7^{+8.7}_{-7.6}$ |

## FIGURES

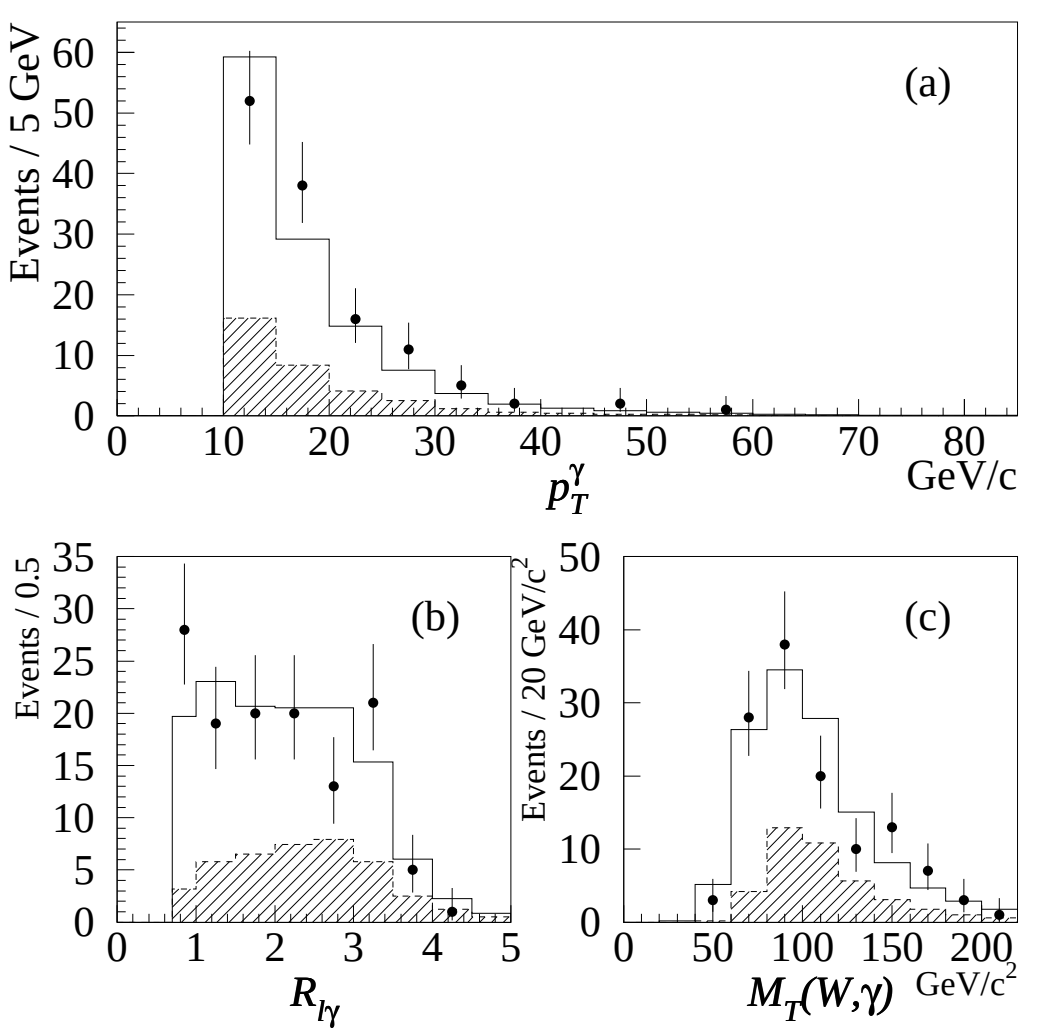

FIG. 1. (a) The $p_T^\gamma$ spectrum for the 127 $W\gamma$ candidates. The $\mathcal{R}_{\ell\gamma}$ and $M_T(W,\gamma)$ distributions are shown in (b) and (c). The solid circles with error bars are the data. The open histogram is the sum of the SM Monte Carlo prediction plus the background estimate (shown as shaded histogram).

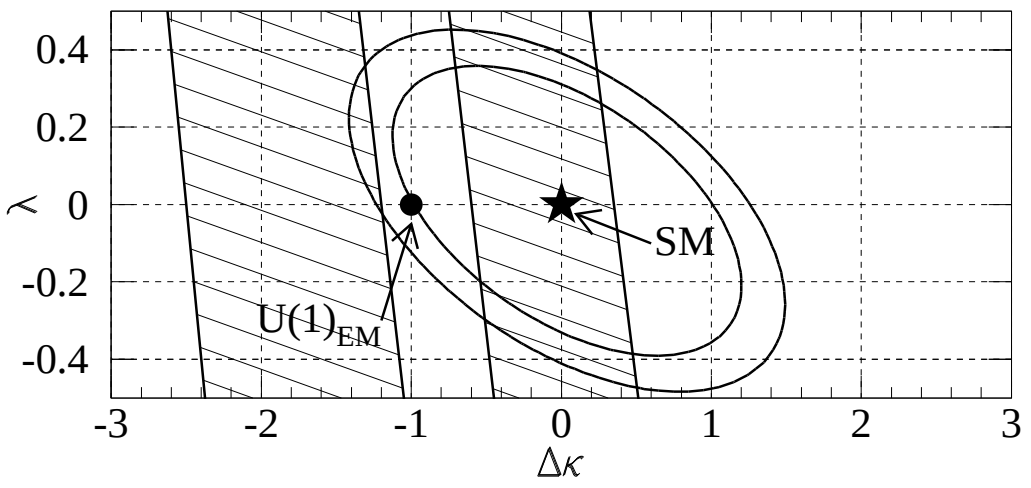

FIG. 2. Limits on the $CP$-conserving anomalous coupling parameters $\Delta\kappa$ and $\lambda$. The inner and outer ellipses represent the one- and two-dimensional 95% CL exclusion contours respectively. The shaded bands represent the regions allowed by the CLEO one-dimensional 95% CL limits [4].